\documentclass[showkeys,showpacs,prd,superscriptaddress]{revtex4}
\usepackage{amsmath}
\usepackage{amssymb}
\usepackage{graphicx}
\usepackage{color}
\usepackage{ulem}
\usepackage{tikz}

\begin{document}

\title{
Extended objects in nonperturbative quantum field theory \\
and the cosmological constant
}

\author{
Vladimir Dzhunushaliev
}
\email{v.dzhunushaliev@gmail.com}
\affiliation{
Dept. Theor. and Nucl. Phys., KazNU, Almaty, 050040, Kazakhstan
}
\affiliation{
IETP, Al-Farabi KazNU, Almaty, 050040, Kazakhstan
}
\author{
Vladimir Folomeev
}
\email{vfolomeev@mail.ru}
\affiliation{
Institute of Physicotechnical Problems and Material Science of the NAS
of the Kyrgyz Republic, 265~a, Chui Street, Bishkek, 720071,  Kyrgyz Republic
}
\author{
Burkhard Kleihaus
}
\email{b.kleihaus@uni-oldenburg.de }
\affiliation{
Institut f\"ur Physik, Universit\"at Oldenburg, Postfach 2503
D-26111 Oldenburg, Germany
}
\author{
Jutta Kunz
}
\email{jutta.kunz@uni-oldenburg.de}
\affiliation{
Institut f\"ur Physik, Universit\"at Oldenburg, Postfach 2503
D-26111 Oldenburg, Germany
}

\begin{abstract}
We consider a gravitating extended object constructed from vacuum fluctuations
of nonperturbatively quantized non-Abelian gauge fields.
An approximate description of such an object is given by two gravitating
scalar fields. The object has a core filled with a constant energy density
of the vacuum fluctuations of the quantum fields.
The core is located inside a cosmological event horizon.
An exact analytical solution of the Einstein equations for such a core
is presented. The value of the energy density of the vacuum fluctuations
is connected with the cosmological constant.
\end{abstract}

\pacs{}
\keywords{cosmological constant; nonperturbative quantum field theory}

\maketitle

\section{Introduction}

Probably the first explanation for the present acceleration of the Universe
came from quantum field theory: here the energy of the
vacuum quantum fluctuations gives rise to a cosmological constant 
and thus a cosmological term in the Einstein equations.
However, this model yields a disagreement of more than 100 orders
of magnitude between the measured value of the cosmological constant
and the theoretical zero-point energy as obtained from
perturbative quantum field theory within the Standard Model,
leading to its description as being
``the worst theoretical prediction in the history of physics" \cite{Hobson}.
In the theoretical procedure employed we would like to emphasize
the word ``perturbative", which
means that the calculation is done by perturbative techniques.
The problem of these perturbative calculations is that without a cutoff
the zero-point fluctuations yield an infinite gravitational energy,
whereas a cutoff at the Planck scale leads to the
Planck energy density which is $\sim 10^{120}$
bigger than the measured cosmological constant.

Currently there are many models for explaining the accelerated
expansion of the Universe: a cosmological constant; quintessence;
modified gravity models; ideas from string theory like brane cosmology;
etc. (For details on this problem see, for example,
the reviews \cite{Padmanabhan:2002ji}-\cite{Sahni:1999gb}.)
Many of these models are based on using some fields which effectively lead to
the acceleration. The essential difference of such models
from the model based on the assumption that the cosmological constant
corresponds to the energy of quantum fluctuations
is that in the first case we must have some field equation(s)
describing such (an) additional field(s),
whereas in the second case some field (gauge field, spinor field)
is in the ground state and quantum fluctuations around this state
lead to the appearing of the cosmological constant.
While this is an advantage for such a point of view,
the calculations in perturbative quantum field theory lead to
a huge discrepancy between this theoretical value and
the observed value of the cosmological constant.

The goal of this paper is to show that the {\rm nonperturbative} vacuum
in quantum field theory has a good chance to yield the cosmological constant.
The idea presented here is as follows.
Let us imagine that we can: {\rm (a) construct a gravitating nonperturbative
model of an extended object in quantum field theory;
(b) calculate the gravitational field created by this object}.
The object itself consists of vacuum fluctuations of non-Abelian gauge fields,
that are present in the Standard Model.
Then the structure of such an object can be as follows:
it possesses a core inside the cosmological event horizon
and a tail outside the horizon.
The core and the tail are filled with vacuum fluctuations.
The boundary between the core and the tail is given by
the cosmological event horizon indicating that we have obtained
a cosmological constant originating from the nonperturbatively quantized fields.

\section{Extended objects from an approximate nonperturbative quantization}
\label{sec2}

Here we follow Refs.~\cite{Dzhunushaliev:2015qls,Dzhunushaliev:2015hoa}
where an approximate nonperturbative quantization procedure was applied
to modeling a glueball. To begin with, let us consider the SU(3) Lagrangian
in quantum chromodynamics
\begin{equation}
\begin{split}
	\mathcal L = &  \frac{1}{4} F^B_{\mu \nu} F^{B \mu \nu} =
	f^a_{\mu \nu} f^{a\mu \nu} +
	f^m_{\mu \nu} f^{m \mu \nu} +
\\	
	&
	g^2 C^{a b_1 c_1} C^{a b_2 c_2} A^{b_1}_\mu A^{c_1}_\nu A^{b_2 \mu} A^{c_2 \nu} +
	g^2 C^{a p_1 q_1} C^{a p_2 q_2} A^{p_1}_\mu A^{q_1}_\nu A^{p_2 \mu} A^{q_2 \nu} +
\\	
	&
	g^2 C^{a b_1 c_1} C^{a p_1 q_1} A^{b_1}_\mu A^{c_1}_\nu A^{p_1 \mu} A^{q_1 \nu} + 	
	\cdots~,
\label{1-10}
\end{split}
\end{equation}
where $B = 1,2, \ldots , 8$ is the SU(3) index;
$F^B_{\mu \nu} = \partial_\mu A^B_\nu - \partial_\nu A^B_\mu +
g C^{BCD} A^C_\mu A^D_\nu$
is the field strength operator;
$f^B_{\mu \nu} = \partial_\mu A^B_\nu -\partial_\nu A^B_\mu $;
$g$ is the coupling constant; $a, b_{1,2}, c_{1,2} = 1,2,3$ are the SU(2) indices;
$m, p_{1,2}, q_{1,2}$ are the coset indices,
and $C^{BCD}$ are the SU(3) structure constants.

In order to obtain an approximation yielding
a nonperturbative description of a glueball, we assume that
\begin{itemize}
\item  expectation values of non-Abelian gauge fields inside a glueball are zero;
\item  2-point Green functions of gauge fields can be approximately represented  through two scalar fields $\phi, \chi$;
\item the behavior of the $SU(2)$ and coset $SU(3)/SU(2)$ components is different; they are described by different scalar fields -- $\phi$ and $\chi$;
\item  4-point Green functions can be approximately decomposed as the product of 2-point Green functions;
\item the expectation value of the product of an odd number of gauge potentials is zero.
\end{itemize}

Thus we assume that it is possible to approximately describe
these quantum fields in the form of two scalar fields,
where one of them (namely $\phi$) describes $SU(2) \in SU(3)$ gauge fields,
and the other one (namely $\chi$) -- the coset $SU(3)/SU(2)$ gauge fields.

\subsection{The effective Lagrangian}

The next step is to obtain some effective Lagrangian for
these two scalar fields. This is done by performing the quantum averaging
of the initial SU(3) Lagrangian.
Here we assume that the 2- and 4-point Green functions
are described in terms of these scalar fields $\phi$ and $\chi$
by using the following relations:
\begin{eqnarray}
	\left\langle
		A^a_\mu A^{a \mu}
	\right\rangle
	& \approx &
	m_1^2 -  \phi^2 ,
\label{1-20}\\
	\left\langle
		\partial_\mu A^a_\alpha \partial^\mu A^{a \alpha}
	\right\rangle
	& \approx &
	 \frac{1}{2} \partial_\mu \phi \partial^\mu \phi ,
\label{1-30}\\
	C_{a b_1 c_1} C_{a b_2 c_2} \left\langle
		A^{b_1}_\mu A^{c_1}_\nu A^{b_2 \mu} A^{c_2 \nu}
	\right\rangle
	&\approx &
	\frac{\lambda_1}{4} \left(
		m_1^2 -  \phi^2
	\right)^2
\label{1-40}\\
	\left\langle
		A^m_\mu A^{m \mu}
	\right\rangle
	&\approx &
	m_2^2 - \chi^2 ,
\label{1-50}\\
	\left\langle
		\partial_\mu A^m_\alpha \partial^\mu A^{m \alpha}
	\right\rangle
	& \approx &
	\frac{1}{2} \partial_\mu \chi \partial^\mu \chi ,
\label{1-60}\\
	C_{a p_1 q_1} C_{a p_2 q_2} \left\langle
		A^{p_1}_\mu A^{q_1}_\nu A^{p_2 \mu} A^{q_2 \nu}
	\right\rangle
	&\approx &
	\frac{\lambda_2}{4} \left(
		m_2^2 -  \chi^2
	\right)^2 ,
\label{1-70}\\
	C_{a b c} C_{a m n}\left\langle
		A^b_\mu A^c_\nu A^{m \mu} A^{n \nu}
	\right\rangle
	&\approx &
	\frac{1}{2} \phi^2 \chi^2  ,
\label{1-80}\\
	\left\langle A^B_\mu \right\rangle = 0 ,
\label{1-90}\\
	\left\langle
		\partial_\mu A^B_\nu  A^C_\rho A^D_\sigma
	\right\rangle = 0 ,
\label{1-100}
\end{eqnarray}
where $a,b,c,d = 1,2,3$ are the SU(2) indices, $m,n,p,q = 4,5,6,7,8$ are the coset indices,
and $\lambda_{1,2}$ and $m_{1,2}$ are closure constants.
Equations \eqref{1-90} and \eqref{1-100} imply that the expectation value
of odd products is zero.
Note here that because of the positivity of the
relations \eqref{1-20}, \eqref{1-50} and \eqref{1-30}, \eqref{1-60}
the squares of the scalar fields $\phi, \chi$ should be less than $m_{1,2}^2$,
respectively.
The effective Lagrangian then becomes
\begin{equation}
	\mathcal L_{\rm{eff}} = \left\langle \mathcal L_{SU(3)} \right\rangle \approx
	\frac{1}{2} \left( \nabla_\mu \phi \right)^2 +
	\frac{1}{2} \left( \nabla_\mu \chi \right)^2 -
	\frac{\lambda_1}{4} \left(
		\phi^2 - m_1^2
	\right)^2  -
	\frac{\lambda_2}{4} \left(
		\chi^2 - m_2^2
	\right)^2  -
	\frac{1}{2}  \phi^2  \chi^2.
\label{1-110}
\end{equation}
The procedure of obtaining  the effective Lagrangian \eqref{1-110}
from the initial Lagrangian \eqref{1-10} is  similar to what happens
in turbulence modeling when one gets the Reynolds equation
starting from the Navier-Stokes equation \cite{Wilcox}.
(A detailed discussion of the similarity between the
nonperturbative quantization and turbulence modeling
is given in Ref.~\cite{Dzhunushaliev:2015hoa}.)

The field equations derived from the Lagrangian \eqref{1-110} are as follows
\begin{eqnarray}
  \partial_\mu \partial^\mu \phi &=&
  - \phi \left[ \chi^2 + \lambda_1
  \left(
    \phi^2 - m_1^2
  \right) \right],
\label{1-120}\\
  \partial_\mu \partial^\mu \chi &=&
  - \chi \left[ \phi^2 + \lambda_2
  \left(
    \chi^2 - m_2^2
  \right) \right].
\label{1-130}
\end{eqnarray}

\subsection{Extended objects}

Before turning to gravitating systems of quantum fluctuations,
let us consider the simpler case of Minkowski spacetime.
In the case of spherical symmetry
and no time-dependence
the equations \eqref{1-120}  and \eqref{1-130}
take the following dimensionless form
\begin{eqnarray}
  {\phi}''+\frac{2}{x}{\phi}' &=& \phi[\chi^2 +
  \lambda_1(\phi^2-m_1^2)],
\label{1-140}\\
  {\chi}''+\frac{2}{x}{\chi}' &=& \chi[\phi^2 +
  \lambda_2(\chi^2 - m_2^2)],
\label{1-150}
\end{eqnarray}
where the dimensionless radial coordinate $x = r/l_0$
has been introduced and
the following redefinitions have been used:
$
\phi \to l_0 \phi $, $\chi \to l_0 \chi , m_{1,2} \to l_0 m_{1,2}$.

We are looking for regular solutions in Minkowski spacetime
whose energy density takes asymptotically a constant value.
An asymptotic expansion of the fields leads to
\begin{eqnarray}
        \phi &\approx& m_1 - \phi_\infty
        \frac{e^{-x \sqrt{2     \lambda_1 m_1^2}}}{x}~,
\label{1-200}\\
        \chi &\approx& \chi_\infty
        \frac{e^{- x \sqrt{ m_2^2 - \lambda_2 m_1^2 }}}{x}~,
\end{eqnarray}
where $\phi$ goes to a finite constant, whereas $\chi$
vanishes asymptotically.

We solve the above set of equations as a nonlinear eigenvalue problem,
where  $m_{1,2}$  are eigenvalues and $\phi, \chi$ are eigenfunctions.
This yields regular solutions with a finite asymptotic energy density.
One can consider this energy density as the nonperturbative energy
density of the vacuum.

\begin{figure}[h]
\begin{minipage}[t]{.45\linewidth}
  \begin{center}
  \includegraphics[width=.95\linewidth]{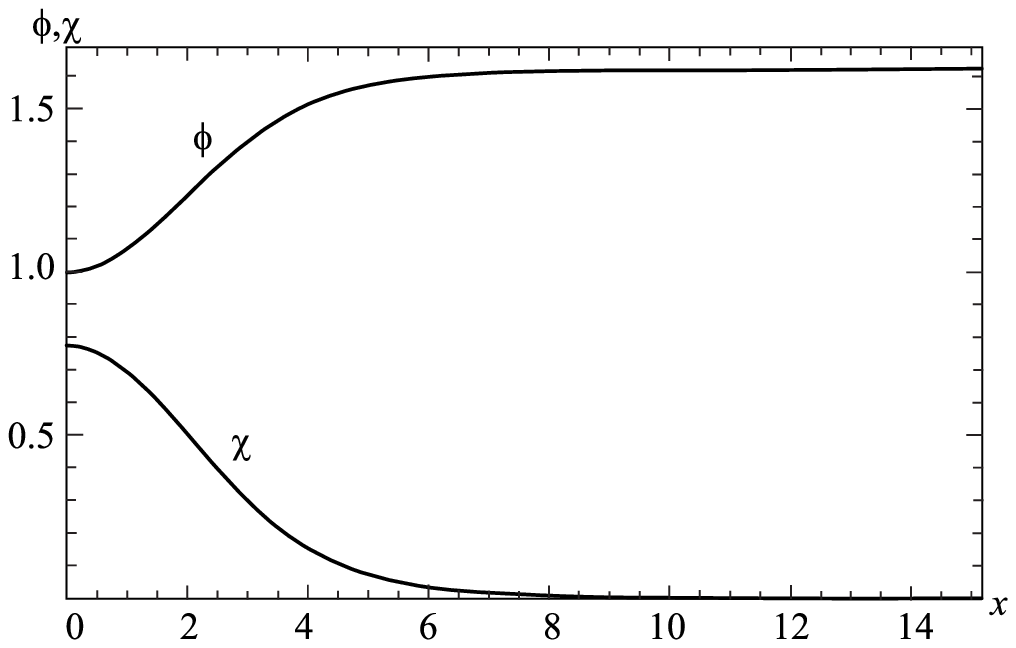}
  \caption{The profiles of $\phi(x)$  and $\chi(x)$ in Minkowski spacetime.
  $\lambda_1 = 0.1; \lambda_2 = 1.0; \phi_0 = 1.0; \chi_0 = \sqrt{0.6};
  m_1 = 1.617168; m_2 = 1.492735$.
  }
  \label{fig1}
  \end{center}
\end{minipage}\hfill
\begin{minipage}[t]{.45\linewidth}
  \begin{center}
  \includegraphics[width=.95\linewidth]{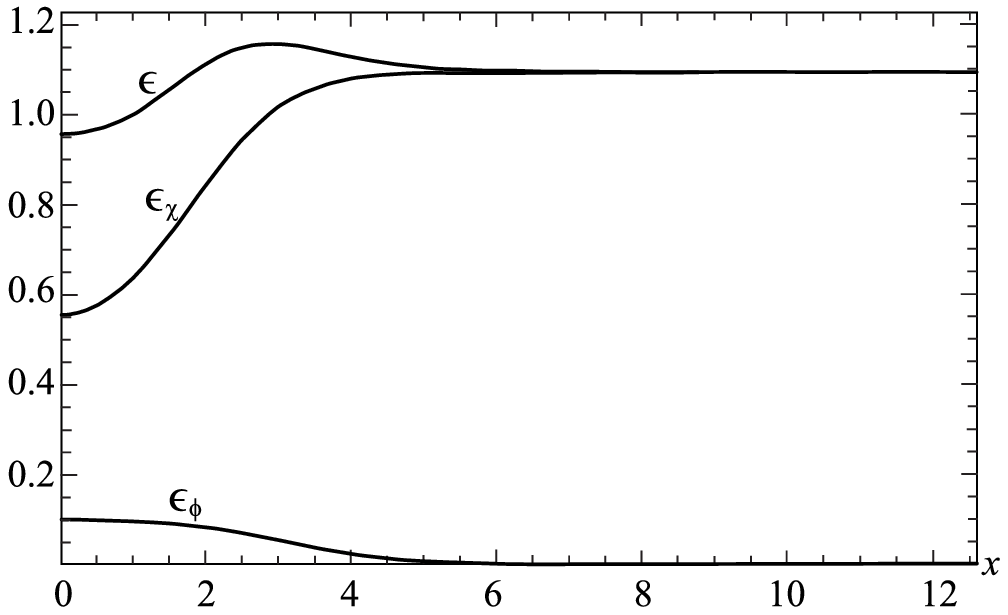}
  \caption{The energy density profiles: for the system as a whole $\epsilon(x)$ (solid line) and for the scalar fields $\epsilon_\phi(x)$   and $\epsilon_\chi (x)$  (dashed lines) in Minkowski spacetime.
   $\lambda_1 = 0.1; \lambda_2 = 1.0; \phi_0 = 1.0; \chi_0 = \sqrt{0.6};
  m_1 = 1.617168; m_2 = 1.492735$.
  }
  \label{fig2}
  \end{center}
\end{minipage}\hfill
\end{figure}

Typical profiles of the functions $\phi(x), \chi(x)$ are presented
in Fig.~\ref{fig1}. In turn, Fig.~\ref{fig2}
shows the energy density profiles for the system as a whole, $\epsilon(x)$,
and for the fields $\phi$ and $\chi$, separately:
\begin{eqnarray}
	\epsilon &=& \frac{1}{2} {\phi'}^2(x) + \frac{1}{2} {\chi'}^2(x) +
	\frac{\lambda_1}{4} \left(
		\phi^2 - m_1^2
	\right) ^2 + \frac{\lambda_2}{4} \left(
		\chi^2 - m_2^2
	\right) ^2 + \frac{1}{2} \phi^2 \chi^2 ,
\label{1-170}\\
	\epsilon_\phi &=& \frac{1}{2} {\phi'}^2(x) +
	\frac{\lambda_1}{4} \left(
		\phi^2 - m_1^2
	\right) ^2 ,
\label{1-180}\\
	\epsilon_\chi &=& \frac{1}{2} {\chi'}^2(x) +
	\frac{\lambda_2}{4} \left(
		\chi^2 - m_2^2
	\right) ^2~.
\label{1-190}
\end{eqnarray}

The asymptotic behavior of the dimensionless energy densities is as follows
\begin{eqnarray}
	\epsilon &\approx& 2 \lambda_1 m_1^2
	\frac{e^{-2 x \sqrt{2 \lambda_1 m_1^2}}}{x^2} +
	\frac{1}{2} \chi_\infty^2 \left(
		m_2^2 - \lambda_2 m_1^2
	\right)
	\frac{e^{-2 x \sqrt{m_2^2 - \lambda_2 m_1^2}}}{x^2}
\nonumber \\
	&&
	+\frac{1}{2} m_1^2 \chi_\infty^2
	\frac{e^{-2 x \sqrt{m_2^2 - \lambda_2 m_1^2}}}{x^2} +
	\frac{\lambda_2}{4} m_2^4 ,
\label{1-220}\\
	\epsilon_\phi &\approx& 2 \lambda_1 m_1^2
	\frac{e^{-2 x \sqrt{2 \lambda_1 m_1^2}}}{x^2} ,
\label{1-230}\\
	\epsilon_\chi &\approx& \frac{1}{2} \chi_\infty^2 \left(
		m_2^2 - \lambda_2 m_1^2
	\right)
	\frac{e^{-2 x \sqrt{m_2^2 - \lambda_2 m_1^2}}}{x^2} +
	\frac{\lambda_2}{4} m_2^4 .
\label{1-240}
\end{eqnarray}

Thus this extended object is constructed in such a way
that the quantum fluctuations of the non-Abelian gauge field $A^m_\mu$
(described here by the scalar field $\chi$)
displace the quantum fluctuations of the non-Abelian gauge field $A^a_\mu$
(described here by the scalar field $\phi$).
This effect is similar to the Meissner effect in superconductivity
when a magnetic field is expelled from the superconductor
during its transition to the superconducting state.

According to \eqref{1-220}, the vacuum energy density (with the dimension cm$^{-4}$) is approximately
equal to
$(\lambda_2/4) m_2^4$ cm$^{-4}$, where we went back to the dimensional quantity
$m_2$ cm$^{-1}$.

\section{Gravitating extended objects}

Now we wish to consider gravitating scalar fields describing
gravitating quantum fluctuations of non-Abelian gauge fields.
Here one may expect that the presence of the density of vacuum fluctuations
gives rise to the appearance of a cosmological event horizon.

To obtain the Lagrangian,
we couple the effective fields $\phi$ and $\chi$ to Einstein gravity.
Thus the Lagrangian consists of the Einstein Lagrangian
and the effective Lagrangian from Eq.~(\ref{1-110})
\begin{equation}
\label{2-10}
  L = -\frac{R}{16\pi G} +  \left[
      \frac{1}{2}\partial_\mu \varphi \partial^\mu
        \varphi + \frac{1}{2}\partial_\mu \chi \partial^\mu
        \chi - V(\varphi,\chi)
    \right]
\end{equation}
with
\begin{equation}
\label{2-20}
   V(\varphi,\chi) =\frac{\lambda_1}{4} \left(
		\phi^2 - m_1^2
	\right)^2  +
	\frac{\lambda_2}{4} \left(
		\chi^2 - m_2^2
	\right)^2  +
	\frac{1}{2}  \phi^2  \chi^2 .
\end{equation}

The corresponding field equations are
\begin{eqnarray}
\label{2-40}
  R_\mu^\nu - \frac{1}{2} \delta_\mu^\nu R &=& \varkappa T_\mu^\nu ,
\\
\label{2-50}
   \square \phi  &=& - \frac{\partial V(\phi, \chi)}{\partial \phi} ,
\\
\label{2-60}
   \square \chi  &=& - \frac{\partial V(\phi, \chi)}{\partial \chi} ,
\end{eqnarray}
where $\varkappa=8\pi G$, and $T_\mu^\nu$ is the energy-momentum tensor
\begin{equation}
    T_\mu^\nu =
        \partial_\mu \phi \partial^\nu  \phi +
        \partial_\mu \chi \partial^\nu \chi -
        \delta_\mu^\nu \left[
            \frac{1}{2}\partial_\alpha \phi \partial^\alpha \phi +
            \frac{1}{2}\partial_\alpha \chi \partial^\alpha \chi -
            V(\phi,\chi)
        \right] .
\label{2-30}
\end{equation}

\subsection{Spherically symmetric Ansatz}

Again we are looking for static spherically symmetric solutions.
Therefore we adopt for the metric the Ansatz
\begin{equation}
	ds^2 = A(r) e^{2 \alpha(r)} dt^2 - \frac{dr^2}{A(r)} -
	r^2 \left(
		d \theta^2 + \sin^2 \theta d \varphi^2
	\right) .
\label{2-70}
\end{equation}
Also, the fields $\phi$ and $\chi$ again depend only on the radial coordinate.

Substituting this Ansatz into the equations \eqref{2-40}-\eqref{2-30}, we find
\begin{eqnarray}
	- A\alpha'' - \frac{A''}{2} - A{\alpha'}^2 - \frac{3}{2} A' \alpha' -
	\frac{A \alpha'}{x} - \frac{A'}{x}
	&=&  \tilde \varkappa  \left[
		\frac{A}{2} \left(
			{\tilde \phi}^{\prime 2} + {\tilde \chi}^{\prime 2}
	\right) + V \left( \tilde \phi, \tilde \chi \right)
	\right]  ,
\label{2-80}\\
	\alpha^\prime &=& \tilde \varkappa \frac{x}{2} \left(
		{\tilde \phi}^{\prime 2} + {\tilde \chi}^{\prime 2}
	\right) ,
\label{2-90}\\
	A' - \frac{1-A}{x} &=&  - \tilde \varkappa x \left[
		\frac{A}{2} \left(
			{\tilde \phi}^{\prime 2} + {\tilde \chi}^{\prime 2}
	\right) + V \left( \tilde \phi, \tilde \chi \right)
	\right]  ,
\label{2-100}\\
	{\tilde \phi}^{\prime \prime} + \left(
		\frac{2}{x} + \alpha^\prime + \frac{A^\prime}{A}
	\right) {\tilde \phi}^\prime  &=&  \frac{\tilde \phi}{A} \left[
		{\tilde \chi}^2 + \lambda_1 \left(
			{\tilde \phi}^2 - {\tilde m}_1^2
		\right)
	\right] ,
\label{2-120}\\
	{\tilde \chi}^{\prime \prime} + \left(
		\frac{2}{x} + \alpha^\prime + \frac{A^\prime}{A}
	\right) {\tilde \chi}^\prime  &=&  \frac{\tilde \chi}{A} \left[
		{\tilde \phi}^2 + \lambda_2 \left(
			{\tilde \chi}^2 - {\tilde m}_2^2
		\right)
	\right] .
\label{2-130}
\end{eqnarray}
Here we have introduced the following dimensionless quantities:
$\tilde \phi = l_0 \phi$,
$\tilde \chi = l_0 \chi$, $\tilde \varkappa = \varkappa/l_0^2$,
${\tilde m}_{1,2} = l_0 m_{1,2}$, $x = r/l_0$,
and $l_0$ is some characteristic length.
(We will see below that this length should be identified
with the radius of the cosmological event horizon.)

\boldmath
\subsection{Taylor expansion of the solutions at the points $x = 0, x_H, \infty$}
\unboldmath

We wish to find solutions which possess a cosmological event horizon.
The presence of the latter implies that  there will be a
point $x_H = r_H/l_0$ at which
\begin{equation}
	A\left(  x_H \right)  = 0 .
\label{2-140}
\end{equation}
Again we aim at solutions where the scalar fields
tend to limiting values for $x\rightarrow \infty$,
with $\tilde \phi \rightarrow \tilde m_1$ and $\tilde \chi \rightarrow 0$.
The asymptotic geometry would, however, now be of de Sitter type.

In order to find such solutions,
let us first consider their Taylor expansions
at the origin, $x = 0$,
at the cosmological horizon, $x = x_H$, and at  infinity, $x \to  \infty$.

\subsubsection{Taylor expansion at the origin}

We assume that at the origin the solutions behave as
\begin{eqnarray}
	A(x) &=& 1 
        + A_2 \frac{x^2}{2} + \ldots ,
\label{2-1-10}\\
	\alpha(x) &=& \alpha_0 + \alpha_2 \frac{x^2}{2} + \ldots ,
\label{2-1-20}\\
	\tilde\phi(x) &=& \tilde\phi_0 + \tilde\phi_2 \frac{x^2}{2} + \ldots ,
\label{2-1-30}\\
	\tilde\chi(x) &=& \tilde\chi_0 + \tilde\chi_2 \frac{x^2}{2} + \ldots .
\label{2-1-40}
\end{eqnarray}
This implies for the center the boundary conditions
\begin{equation}
	A(0) = 1, \quad 
        \tilde\phi'(0) = \tilde\chi'(0) = 0.
\label{2-1-50}
\end{equation}


\subsubsection{Taylor expansion at the cosmological event horizon}
\label{TaylorCEH}

At the cosmological event horizon the solutions should satisfy
\begin{eqnarray}
	A(x) &=& A'_H \left( x - x_H \right) + \ldots ,
\label{2-2-10}\\
	\alpha(x) &=& \alpha_H + \ldots ,
\label{2-2-20}\\
	\tilde\phi(x) &=& \tilde\phi_H +\tilde \phi'_H \left( x - x_H \right) + \ldots ,
\label{2-2-30}\\
	\tilde\chi(x) &=& \tilde\chi_H +\tilde \chi'_H \left( x - x_H \right) + \ldots .
\label{2-2-40}
\end{eqnarray}
From Eqs.~\eqref{2-120} and \eqref{2-130}, one can find
\begin{eqnarray}
	\tilde\phi' \left(  x_H \right)  &=& \tilde\phi'_H =
	\frac{\tilde\phi_H}{A'_H} \left[
		{\tilde \chi}^2_H + \lambda_1 \left(
			{\tilde \phi}^2_H - {\tilde m}_1^2
		\right)
	\right],
\label{2-2-50}\\
	\tilde\chi' \left(  x_H \right)  &=& \tilde\chi'_H =
	\frac{\tilde\chi_H}{A'_H} \left[
		{\tilde \phi}^2_H + \lambda_2 \left(
			{\tilde \chi}^2_H - {\tilde m}_2^2
		\right)
	\right],
\label{2-2-60}
\end{eqnarray}
where $A'_H$ is obtained from \eqref{2-100} as
\begin{equation}
	A'_H = \frac{1}{x_H} - \tilde \varkappa \, x_H
		 V \left( {\tilde \phi}_H, {\tilde \chi}_H \right).
\label{2-2-70}
\end{equation}

\subsubsection{Taylor expansion at infinity}

At infinity the solutions possess the asymptotic behavior
\begin{eqnarray}
	A(x) &\approx& -A_\infty x^2 ,
\label{2-210}\\
	\tilde\phi(x) &\approx& \tilde m_1 + \tilde\phi_\infty \left(x^{\beta_1}+x^{\beta_2}  \right),
	\quad
	\beta_{1,2} = - \frac{3}{2} \pm
	\sqrt{\frac{9}{4} - \frac{2 m_1^2 \lambda_1}{A_\infty}} ,
\label{2-220}\\
	\tilde\chi(x) &\approx& \tilde\chi_\infty \left(x^{\gamma_1}+x^{\gamma_2}  \right),
	\quad
	\gamma_{1,2} = - \frac{3}{2} \pm
	\sqrt{\frac{9}{4} - \frac{m_1^2 - \lambda_2 m_2^2 }{A_\infty}} ,
\label{2-230}
\end{eqnarray}
where $ \tilde\phi_\infty$ and  $\tilde\chi_\infty$ are integration constants.
Using these expressions, the asymptotic form for $\alpha(x)$
can be found from Eq.~\eqref{2-90}.

Depending on the sign of the expressions under the square roots
in $\beta_{1,2}$ and $\gamma_{1,2}$, the asymptotic behavior
of the scalar fields changes drastically.
For positive values of the expressions under the square roots,
a power damping will take place.
For negative values of the expressions under the square roots, however,
 Eqs.~\eqref{2-220} and \eqref{2-230} take the following form
\begin{eqnarray}
\label{phi_asym_osc}
\tilde\phi(x) &\approx& \tilde m_1 +
2 \tilde\phi_\infty	x^{-3/2}\cos{\left(\sqrt{\Big |\frac{9}{4} - \frac{2 m_1^2 \lambda_1}{A_\infty}\Big |}\ln{x}\right)},\\
\tilde\chi(x) &\approx&
2 \tilde\chi_\infty	x^{-3/2}\cos{\left(\sqrt{\Big |\frac{9}{4} - \frac{m_1^2 - \lambda_2 m_2^2 }{A_\infty}\Big |}\ln{x}\right)}.
\label{chi_asym_osc}
\end{eqnarray}

\subsection{Numerical solutions}
\label{numcalc}

We solve the system of equations  \eqref{2-90}-\eqref{2-130} numerically.
As discussed above, we seek solutions
possessing a cosmological event horizon.

Our strategy for obtaining such solutions is to employ a two-step procedure.
We first solve the equations in the inner region $x<x_H$.
Subsequently, we integrate the equations in the outer region $x> x_H$.
We solve the set of equations  \eqref{2-90}-\eqref{2-130}
as a nonlinear eigenvalue problem with eigenvalues $m_{1,2}$
and $\tilde \varkappa$ and eigenfunctions $\tilde \phi, \tilde \chi$
and $A, \alpha$.
For the numerical computations, we employ the shooting method.
In the first step we
start the solution at the point $x_H - \delta$
with $\delta = 10^{-3}$, and integrate towards the center of the system,
i.e., in the direction $x < x_H$.
As boundary conditions, we choose a set of values
$A_H=0$, $\alpha_H$, $\tilde \phi_H$ and $\tilde \chi_H$ at
the cosmological horizon, and then determine the unknown
values $m_{1,2}$ and $\tilde \varkappa$,
by requiring the boundary conditions \eqref{2-1-50}.

Once a numerical solution inside the cosmological event horizon is found,
we seek the solution outside the event horizon, $x >x_H$.
To do this, we solve Eqs.~\eqref{2-90}-\eqref{2-130} with the values
of the parameters $m_{1,2}$ and $\tilde \varkappa$
found in the region with $x < x_H$.

\begin{figure}[h]
\begin{minipage}[t]{.45\linewidth}
  \begin{center}
  \includegraphics[width=.95\linewidth]{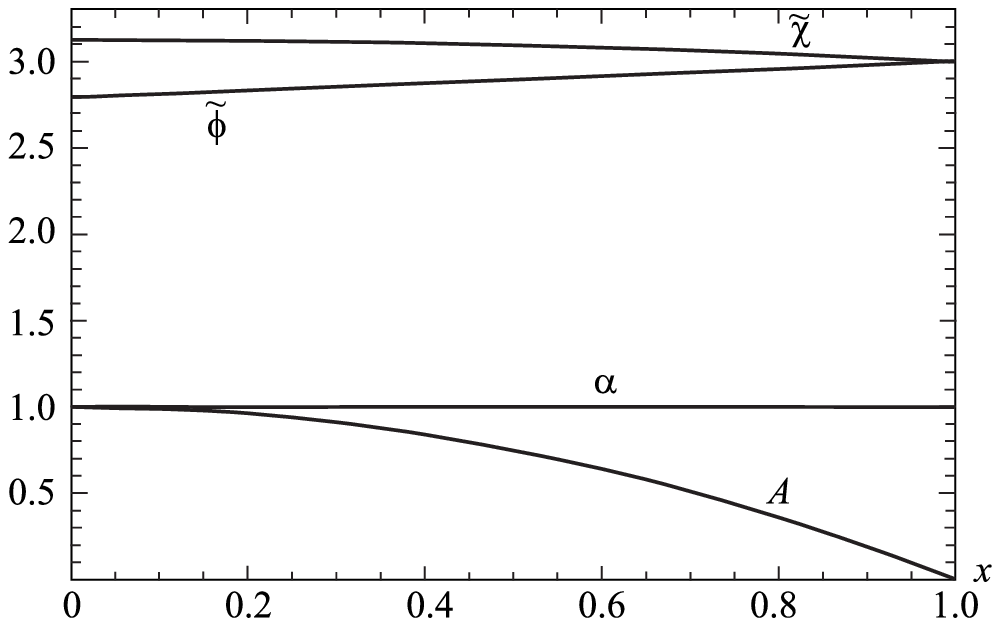}
  \caption{Solution inside the cosmological event horizon.}
  \label{fig3}
  \end{center}
\end{minipage}\hfill
\begin{minipage}[t]{.45\linewidth}
  \begin{center}
  \includegraphics[width=.95\linewidth]{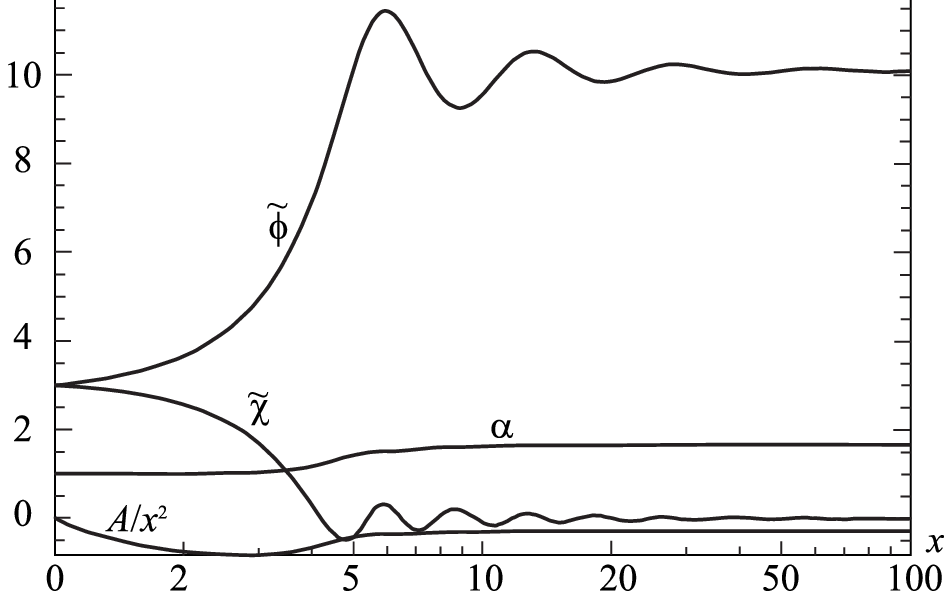}
  \caption{Solution outside the cosmological event horizon,
where asymptotically $\tilde \phi \to \tilde m_1$,
  $\tilde \chi \to 0$.}
  \label{fig4}
  \end{center}
\end{minipage}\hfill
\end{figure}

Figs.~\ref{fig3} and \ref{fig4} show an example of such solutions
for the following parameters and horizon values of the fields:
\begin{equation}
	\lambda_1 = 0.1, \quad \lambda_2 = 1.0, \quad
        x_H=1.0, \quad \alpha_H= 1.0 , \quad
	\tilde \phi_H = 3.0, \quad \tilde \chi_H = 3.0 ,
\label{2-180}
\end{equation}
and the associated eigenvalues
\begin{equation}
	\tilde m_1 = 10.07462, \quad \tilde m_2 = 4.2241352, \quad
	\tilde \varkappa = 0.01095.
\label{2-190}
\end{equation}
This solution exhibits the oscillating behavior
according to the asymptotic expansions \eqref{phi_asym_osc} and \eqref{chi_asym_osc}.

As one can see in Fig.~\ref{fig3},
inside the cosmological event horizon the scalar fields change very slowly.
Therefore the energy density of the quantum fluctuations
is also almost constant
with the main contribution coming from the potential part.
Since the solutions start at the horizon with the boundary conditions
\eqref{2-2-30} and \eqref{2-2-40},
the energy density approximately corresponds to the cosmological constant:
\begin{equation}
	\Lambda = \frac{1}{l_0^2} {\tilde T}^0_0 =
	\frac{1}{l_0^2} \tilde \varkappa \left[
		\frac{\lambda_1}{4} \left(
		{\tilde \phi}_H^2 - {\tilde m}_1^2
	\right)^2  +
	\frac{\lambda_2}{4} \left(
		{\tilde \chi}_H^2 - {\tilde m}_2^2
	\right)^2  +
	\frac{1}{2}  {\tilde \phi}_H^2  {\tilde \chi}_H^2
	\right]  = \frac{3}{l_0^2} .
\label{4-80}
\end{equation}
If we choose $l_0 \approx 10^{28}$ cm
(i.e., the radius of the observable Universe),
we obtain the expected value of the cosmological constant
$\Lambda \approx 10^{-56} \text{cm}^{-2}$.

The presented numerical solution is given for $l_0 \approx l_{Pl}$,
and it would be very important to show that such a solution does exist for $l_0 \gg l_{Pl}$.
It will be done in the next section.

\subsection{Analytical solution inside the cosmological event horizon}
\label{IIId}

The system of equations \eqref{2-80}-\eqref{2-130}
has a trivial de Sitter solution of the form
\begin{eqnarray}
	\tilde\phi(x) &=& \tilde\phi_H ,
\label{4-10}\\
	\tilde\chi(x) &=&\tilde \chi_H ,
\label{4-20}\\
	A(x) &=& 1 - x^2 ,
\label{4-30}\\
	\alpha(x) &=& \alpha_H
\label{4-40}
\end{eqnarray}
with the following eigenvalues
\begin{eqnarray}
	\tilde m_1 &=& \sqrt{\frac{\tilde \chi_H^2}{\lambda_1} +
		\tilde \phi_H^2	
	} ,
\label{4-50}\\
	\tilde m_2 &=& \sqrt{\frac{\tilde \phi_H^2}{\lambda_2} +
		\tilde \chi_H^2	
	} ,
\label{4-60}\\
	\tilde \varkappa &=& \frac{3}{
		\frac{{\tilde \chi}_H^4}{4 \lambda_1} +
		\frac{{\tilde \phi}_H^4}{4 \lambda_2} +
		\frac{{\tilde \phi}_H^2 {\tilde \chi}_H^2}{2}
	} .
\label{4-70}
\end{eqnarray}
In principle, this solution is valid over all space.

Let us consider the case where $\tilde\varkappa$ in \eqref{4-70} is $\tilde\varkappa \approx 8 \pi l^2_{Pl}/l_0^2$. In this case
\begin{equation}
	\frac{\chi_H^4}{4 \lambda_1} +
	\frac{\phi_H^4}{4 \lambda_2} +
	\frac{\phi_H^2 {\chi}_H^2}{2} \approx
	\frac{3}{8 \pi} \left( \frac{1}{l_0 l_{Pl}} \right)^2 =
	\frac{3}{8 \pi} \frac{\Lambda}{l_{Pl}^2} \approx 10^{9} \mathrm{cm}^{-4}.
\label{4-90}
\end{equation}
For these values of the fields $\phi, \chi$ we have the observed value of the cosmological constant
$\Lambda \approx 10^{-56}$ cm$^{-2}$. These values of the fields give us the following vacuum energy for the nonperturbatively quantized fields:
\begin{equation}
	\epsilon_{vac} \approx \hbar c V\left( \phi_H, \chi_H\right) =
	\hbar c \left(
		\frac{\chi_H^4}{4 \lambda_1} +
		\frac{\phi_H^4}{4 \lambda_2} +
		\frac{\phi_H^2 {\chi}_H^2}{2}
	\right) \approx 10^{-8} \mathrm{erg/cm}^3 ,
\label{4-100}
\end{equation}
i.e. the energy density of the present Universe.
The solution for the fields $\phi$ and $\chi$ from \eqref{4-90} is a very good approximation for the real Universe since
it has a constant distribution of the nonperturbative vacuum energy density for the gauge fields $A^B_\mu$.
It must be remembered  here  that $\phi$ describes the dispersion of nonperturbative quantum fluctuations of the gauge fields $A^a_\mu \in SU(2) \subset SU(3)$ and $\chi$
describes the dispersion of nonperturbative quantum fluctuations of the coset fields $A^m_\mu \in SU(3) / SU(2)$.

\section{Discussion and conclusion}

Here we have shown that a nonperturbatively quantized non-Abelian gauge field
permits the existence of a regular extended object, which
in the presence of gravity may possess a cosmological event horizon.
Inside the  horizon, the energy density of the scalar fields
can be made practically constant, and thus may be (approximately) considered
as a cosmological constant.

The size of such an extended object corresponds to the size of the Universe
as a whole, and its core, the observable Universe, corresponds to the region
located inside the cosmological event horizon.
Let us note that in Section \ref{numcalc} we have obtained a solution with
$$\bar \varkappa = \varkappa / l_0^2 =8\pi l^2_{Pl}/l_0^2 \approx 10^{-2} .$$
This means that $l_0 \approx 50 l_{Pl}$.
In Section \ref{IIId} we have shown that on the event horizon there exist such values  of the fields $\phi_H,\chi_H$  
which give the observable value of the cosmological constant:
$$
l_0 \approx \Lambda^{-1/2},
\quad \quad \bar \varkappa \approx l^2_{Pl} \Lambda \approx 10^{-123} .
$$
In this case an extended object created by two scalar fields $\phi, \chi$
(describing the nonperturbatively quantized non-Abelian SU(3) gauge field)
will be immersed into the Universe with the observed value of the
cosmological constant.

The main goal of the paper was to show that
(in contrast to perturbative quantization)
nonperturbative quantization leads to a finite
vacuum energy density, which can be regarded as the cosmological constant.
The main difference compared to perturbative quantization is thus the finiteness
of the vacuum energy density.


\section*{Acknowledgements}

We gratefully acknowledge support provided by the Volkswagen Foundation.
This work was further supported
by the Grant $\Phi.0755$ in fundamental research in natural sciences
by the Ministry of Education and Science of Kazakhstan,
by the DFG Research Training Group 1620 ``Models of Gravity'',
and by FP7, Marie Curie Actions, People,
International Research Staff Exchange Scheme (IRSES-606096).

\end{document}